\documentclass[onecolumn,secnumarabic,amssymb, nobibnotes, aps,notitlepage, pr,superscriptaddress,10pt]{revtex4-2}

\usepackage{amsmath,amssymb}
\usepackage{stackengine,graphicx}
    \graphicspath{{images/}} 
\usepackage{natbib}
\usepackage[dvipsnames]{xcolor}
\usepackage{bbm}
\usepackage{booktabs}
\usepackage{graphicx}
\usepackage{enumitem}
\usepackage{colortbl}
\usepackage{lipsum}
\usepackage{rotating}

\usepackage{framed}

\usepackage{tabu}
\usepackage{makecell}
\usepackage{verbatim}

\usepackage{xurl}

\PassOptionsToPackage{hyphens}{url}\usepackage{hyperref}

\usepackage{algorithm}
\usepackage{algpseudocode}
\usepackage{soul}

\usepackage{collectbox}
\usepackage{bbm}
\usepackage{capt-of}

\UseRawInputEncoding
\usepackage[utf8]{inputenc}

\usepackage{longtable}

\usepackage{hyperref}
\hypersetup{
    colorlinks=true,
    linkcolor=blue,
    filecolor=magenta,      
    urlcolor=cyan,
    citecolor=black
}

\def\l{\left}
\def\r{\right}

\usepackage{array}
\newcolumntype{L}[1]{>{\raggedright\let\newline\\\arraybackslash\hspace{0pt}}m{#1}}
\newcolumntype{C}[1]{>{\centering\let\newline\\\arraybackslash\hspace{0pt}}m{#1}}
\newcolumntype{R}[1]{>{\raggedleft\let\newline\\\arraybackslash\hspace{0pt}}m{#1}}

\newcommand{\f}{\frac}

\usepackage{array}
\newcolumntype{L}[1]{>{\raggedright\let\newline\\\arraybackslash\hspace{0pt}}m{#1}}
\newcolumntype{C}[1]{>{\centering\let\newline\\\arraybackslash\hspace{0pt}}m{#1}}
\newcolumntype{R}[1]{>{\raggedleft\let\newline\\\arraybackslash\hspace{0pt}}m{#1}}

\usepackage{accents}

\usepackage[linewidth=1pt]{mdframed}

\newcommand{\luPsyc}{Department of Psychology, Lehigh University, Bethlehem, Pennsylvania, United States of America}

\newcommand{\lucoh}{Department of Community and Population Health, College of Health, Lehigh University, Bethlehem, Pennsylvania, United States of America}

\newcommand{\bch}{Boston Children’s Hospital, Boston, MA, United States of America}

\newcommand{\hms}{Harvard Medical School, Boston, MA, United States of America}

\newcommand{\umass}{Dept. of Biostatistics and Epidemiology, School of Public Health and Health Sciences, University of Massachusetts Amherst, Amherst, Massachusetts, United States of America}

\newcommand{\uva}{Biocomplexity Institute and Initiative, University of Virginia, Charlottesville, Virginia, United States of America}

\newcommand{\lehighcivil}{Department of Civil and Environmental Engineering, P.C. Rossin College of Engineering and Applied Science, Lehigh University, Bethlehem, Pennsylvania, United States of America}

\newcommand{\mitt}{MIT Computer Science and Artificial Intelligence Laboratory, Cambridge, Massachusetts, United States of America}

\newcommand{\metaculus}{Metaculus, Santa Cruz, California, United States of America
}
\newcommand{\lshtm}{Centre for Mathematical Modelling of Infectious Diseases, London School of Hygiene \& Tropical Medicine, Keppel St, London, WC1E 7HT, UK}

\newcommand{\cumail}{Department of Epidemiology, Mailman School of Public Health, Columbia University, New York, United States.
}
\newcommand{\emory}{Department of Biostatistics, Rollins School of Public Health, Emory University, Atlanta, Georgia, United States of America
}

\usepackage{xurl}

\date{\today} 

\usepackage[paperwidth=8.5in,paperheight=11.0in,
  left=1.in,right=1.in,top=1.5in,bottom=1.5in,
  includefoot,heightrounded]{geometry}

\begin{document}

\title{Assessing Human Judgment Forecasts in the Rapid Spread of the Mpox Outbreak: Insights and Challenges for Pandemic Preparedness}

\author{Thomas~McAndrew}
\affiliation{\lucoh}
\email{mcandrew@lehigh.edu}

\author{Maimuna S. Majumder}
\affiliation{\bch}
\affiliation{\hms}

\author{Andrew A. Lover}
\affiliation{\umass}

\author{Srini Venkatramanan}
\affiliation{\uva}

\author{Paolo Bocchini}
\affiliation{\lehighcivil}

\author{Tamay Besiroglu}
\affiliation{\mitt}
\affiliation{\metaculus}

\author{Allison Codi}
\affiliation{\emory}

\author{Gaia Dempsey}
\affiliation{\metaculus}

\author{Sam Abbott}
\affiliation{\lshtm}

\author{Sylvain Chevalier}
\affiliation{\metaculus}

\author{Nikos I. Bosse}
\affiliation{\lshtm}
\affiliation{\metaculus}

\author{Juan Cambeiro}
\affiliation{\cumail}
\affiliation{\metaculus}

\author{David Braun}
\affiliation{\luPsyc}

\setlength{\parskip}{0.5em}
\setlength{\parindent}{0pt}

\begin{abstract}
In May 2022, mpox~(formerly monkeypox) spread to non-endemic countries rapidly. 
Human judgment is a forecasting approach that has been sparsely evaluated during the beginning of an outbreak.
We collected---between May 19, 2022 and July 31, 2022---1275 forecasts from 442 individuals of six questions about the mpox outbreak where ground truth data are now available.
Individual human judgment forecasts and an equally weighted ensemble were evaluated, as well as compared to a random walk, autoregressive, and doubling time model.
We found (1) individual human judgment forecasts underestimated outbreak size, (2) the ensemble forecast median moved closer to the ground truth over time but uncertainty around the median did not appreciably decrease, and (3) compared to computational models, for 2-8 week ahead forecasts, the human judgment ensemble outperformed all three models when using median absolute error and weighted interval score; for one week ahead forecasts a random walk outperformed human judgment.
We propose two possible explanations: at the time a forecast was submitted, the mode was correlated with the most recent (and smaller) observation that would eventually determine ground truth. 
Several forecasts were solicited on a logarithmic scale which may have caused humans to generate forecasts with unintended, large uncertainty intervals.
To aide in outbreak preparedness, platforms that solicit human judgment forecasts may wish to assess whether specifying a forecast on logarithmic scale matches an individual's intended forecast, support human judgment by finding cues that are typically used to build forecasts, and, to improve performance, tailor their platform to allow forecasters to assign zero probability to events. 
\end{abstract}

\maketitle

\section{Introduction} 
An outbreak of mpox~(formerly monkeypox) was detected in early May 2022, and on July 23, 2022, the World Health Organization declared this outbreak a Public Health Emergency of International Concern (PHEIC)~\cite{whompox,kroger2023mpox,laurenson2023description,van2023mpox,li2023clinical,mellou2023overview,besombes2023investigation,wolfe2023use,kroger2023mpox}.
To date, there have been 87k cases of human mpox reported by at least 110 countries in association with this outbreak~\cite{cdcmpox,aden2023monkeypox}.
Transmission has occurred most frequently between men who have sex with men (MSM); however, cases are not restricted to this population~\cite{endo2022heavy,kupferschmidt2022monkeypox,allan2023position}.

Historically, forecasts of public health metrics of interest---including reported incident cases, deaths, and hospitalizations---have been used to support decision making during outbreaks~\cite{lutz2019applying,george2019technology,kobres2019systematic,ray2020ensemble,viboud2019future}.
Agencies such as the Centers for Disease Control and Prevention (CDC) have routinely hosted forecasting challenges, where they elicit forecasts of public health metrics from research groups during outbreaks of communicable diseases~\cite{biggerstaff2018results,mcgowan2019collaborative,lutz2019applying,cdcmpoxtecnreport}.
These agencies have invested considerable time and resources to host platforms such as the Epidemic Prediction Initiative~\cite{cdcepi} that are designed to collect and communicate probabilistic forecasts of infectious diseases such as seasonal influenza, dengue, neuroinvasive West-Nile Virus disease, Zika virus disease, and COVID-19~\cite{biggerstaff2018results,mcgowan2019collaborative,del2018summary,holcomb2023evaluation}. 

Computational models typically train on existing surveillance data and other external data sources that are hypothesized to be correlated with the forecasting target of interest~\cite{teng2017dynamic,mcgough2017forecasting,hii2012forecast,kiang2021incorporating,osthus2022fast,gibson2020real,ben2019forecasting,turtle2021accurate,borchering2021modeling,borchering2021modeling,maziarz2020agent,becker2021development,howerton2021synergistic}.
Models may also leverage interventions or behavioral adaptations that could influence future disease dynamics~\cite{lemaitre2021scenario,runge2023scenario,heesterbeek2015modeling}. 
Computational models may be either statistical (i.e., leveraging associations between a set of variables and epidemiological outcomes), mechanistic (i.e., encoding domain knowledge on the known causal linkages between population and disease spread dynamics), or a hybrid of both approaches~\cite{haber1988models,becker1999statistical,tang2020review,siettos2013mathematical,metcalf2017opportunities}.
Past work has shown that statistical and mechanistic models are able to incorporate a diverse range of data sources external to standard health surveillance systems to produce more accurate forecasts.
For example, these models have included Google searches for symptoms associated with disease or social media posts related to a disease, genomics data, human mobility data, and meteorological data ~\cite{teng2017dynamic,mcgough2017forecasting,hii2012forecast,kiang2021incorporating,majumder2016utilizing,aiken2020real,poirier2020role}. 
Furthermore, mechanistic models in particular have had past success in describing how the trajectory of an infectious agent may change after choosing one of many potential interventions~\cite{osthus2022fast,gibson2020real,ben2019forecasting,turtle2021accurate,borchering2021modeling}.
Because mechanistic models assume a set of rules for how an infection spreads throughout a population, these models can also be useful for interpreting surveillance data and testing potential public health interventions~\cite{borchering2021modeling,maziarz2020agent,becker2021development,howerton2021synergistic}.

In contrast to statistical and mechanistic forecasts, human judgment forecasts collect and combine predictive densities from a set of individuals~\cite{mcandrew2022chimeric,allipaper,bosse2022comparing,farrow2017human,venkatramanan2022utility,recchia2021well,braun2022crowdsourced}. 
Humans may decide to submit a prediction that was produced by a computational model, from their intuition, or both. 
Past work using human judgment to forecast infectious diseases has shown that human judgment forecasts can perform similarly or at times better compared to computational models~\cite{mcandrew2022chimeric,allipaper,bosse2022comparing,farrow2017human}, can often predict a broader class of targets than computational models~\cite{mcandrew2022expert,mcandrew2022early}, can add data that complements traditional surveillance data~(i.e.~case counts)~\cite{braun2022crowdsourced,mcandrew2024chimeric}, and may perform better than computational models when data on a pathogen is sparse~\cite{recchia2021well,mcandrew2022aggregating,mcandrew2021aggregating,venkatramanan2022utility}.

Within two weeks of the first reported case of mpox in London, six questions---created by the authors of this study---were posted to a community of more than 15,000 registered forecasters on a public prediction platform, Metaculus~\cite{metaculus}. 
During the study period, from May 19, 2022 to July, 31, 2022, the goal of these six questions was to understand the potential intensity and burden of mpox. 
These questions were:  (i) what will be the total number of confirmed and suspected monkeypox cases in Europe as of July 1, 2022?, (ii) what will be the total number of confirmed and suspected monkeypox cases in the United States as of July 1, 2022?, (iii) what will be the total number of confirmed and suspected monkeypox cases in Canada as of July 1, 2022? (iv) how many countries will confirm at least one case of monkeypox by July 31, 2022?, (v) how many states in the United States will report having a case of monkeypox as of July 1, 2022?, and (vi) will the World Health Organization declare the spread of monkeypox a Public Health Emergency of International Concern before 2023? We received a total of 1275 probabilistic predictions (including initial forecasts and subsequent updates) from 442 individual forecasters. 

Here, we evaluate the performance of a human judgement ensemble~(HJE) of the 2022 mpox outbreak, as well as the individual human judgment forecasts that comprised the ensemble. 
We show that the majority of individuals underestimated the size of the outbreak and that, though the median prediction was accurate, the level of uncertainty remained the same or did not appreciably decrease as the time between when the forecast was generated and the ground truth~(for example the total number of cases for a country) decreased.
We propose two potential mechanisms for these observations: (1)~for each question, the most recent observation from the data source that would determine the final ground truth was associated with a given individual's mode prediction and (2)~forecasts were solicited on a logarithmic scale, which may have led to unexpectedly large prediction intervals.

\section{Methods}

\subsection{Collecting crowdsourced predictions}
From May 19, 2022 to July 31, 2022, we collected 1275 individual estimates of six questions posed on the Metaculus platform from 442 users.
Users of Metaculus were permitted to submit a prediction to one or more questions related to mpox. 
Forecasters were not explicitly recruited for the study. 
They were either existing or new users on the Metaculus platform, all of whom have agreed to the terms and conditions outlined by Metaculus~(\url{https://www.metaculus.com/terms-of-use/}).
Members of the Metaculus team advertised mpox questions on their mailing list, which is sent to all Metaculus users, and on Twitter.
We also sent personal emails to those who were considered experts in the modeling of infectious diseases, asking if they would volunteer their time and submit predictions. However, only one such individual identified themselves as submitting predictions~\cite{mcandrew2022early}. 

The format for how a human submits a forecast on Metaculus can be found in the supplement under the header `Forecast submission'. 
Humans were able to submit a forecast at any time before the date on which the ground truth would be resolved, and they were allowed to revise their original forecast as many times as they wished.
Notably, some questions solicited forecasts on a logarithmic scale while other questions were solicited on natural scale~(see Table~\ref{table.comp_model_data_sources}).

\subsection{Surveillance data}

Surveillance data was used to train computational models that served as control against human judgment forecasts.
Surveillance data is defined as any data source that (1)~updates over time and (2)~contains the ground truth that answers a forecasting task.
Surveillance data was used as ground truth for all the questions posed on Metaculus.
For this work, we found surveillance data that reported over time the number of total cases of mpox; and number countries/states with at least one positive mpox case.
Data sources used to define the ground truth were at a weekly temporal scale and were collected from the Centers for Disease Control and Prevention, Global Health~\cite{global_health} and the European Centre for Disease Control and Prevention~\cite{ecdc}.
Supplemental Table~\ref{table.comp_model_data_sources} provides information about surveillance data sources used to train models.


\subsection{Computational models}
We compared our human judgment ensemble forecasts to (i)~a random walk model~(constant trend), (ii) an autoretrogressive model~(linear trend), and (iii) a doubling-time model~(simple exponential growth). 
These three simple computational models were chosen as a baseline to compare against the ensemble of human judgments.
These models are not an attempt to rigorously predict the mpox outcomes that were posed to human forecasters.
Instead, these three models represent three potential schema for how humans may produce predictions: (1) the random walk supposes a constant trend from the last observed data point and the uncertainty grows with the time between this last observation and the target, thus representing a scenario where a human cannot justify an increasing or decreasing trend; (2) the autoretrogressive model allows for an increasing/decreasing trend that is linear, acknowledging that linear predictions are often used to characterize human predictions~\cite{armstrong2001judgmental}; and (3) the doubling-time allows for exponential increases/decreases in trend.
Notably, past work has shown that humans poorly understand exponential growth and thus is the least likely model employed by humans in this experiment~\cite{melnik2023my,wagenaar1975misperception,heckler2013student}.

Computational models were trained retrospectively, and did not train on data that may have been revised as the mpox outbreak was ongoing.
Models were trained every week from the start of data collection up until 1-8 weeks before the date of ground truth. 
For example, if the ground truth was evaluated at week $T$ then we cut training data for all computational models at $T-1$, $T-2$ weeks etc. 

All three models were fit with a No-U-Turn Sampler~(i.e. Hamiltonian MCMC) using  Numpyro~\cite{phan2019composable}.
For each model, we requested 4,000 warm-up samples and 10,000 draws from a single chain.

For this analysis, note that the question that asked whether the WHO would declare a Public Health Emergency of International Concern~(PHEIC) is not included.
This is because there is no clear mapping from the three computational models to a declaration of PHEIC.

Our random walk model assumes that our time series of interest follows a Gaussian random walk
\begin{align*}
    \sigma^{2} \sim \text{Gamma}(1,1); \; 
    Y_{t} = \sum_{t=0}^{t} \mathcal{N}(0,\sigma^{2})
\end{align*}
where $\mathcal{N}(0,\sigma^{2})$ is a normally distributed random variable with expected value $0$, variance $\sigma^{2}$.
Given training data up until time $T-X$, a forecast at time $T$ was generated as 
\begin{align*}
    Y_{T}  = y_{T-X} + \sum_{t=1}^{X} \mathcal{N}(0,\sigma^{2})
\end{align*}
were $y_{T-X}$ is the last observed value in the training data.

Our autoregressive model~(AR) assumes that our outcome of interest $y_{t}$ is drawn from a sequence of random variables $Y_{t}$ that follow
\begin{align*}
    &\beta \sim \text{Gamma}(1,1) ; \; \sigma^{2} \sim \text{Gamma}(1,1); \;  
    Y_{t}|y_{t-1} \sim \mathcal{N}(\beta y_{t-1}, \sigma^{2})
\end{align*}

Our doubling-time model assumes that our target $y_{1}, y_{2}, \cdots, y_{T}$ follows  
\begin{align*}
    \alpha &\sim \text{Gamma}(1,1); \; \beta \sim \text{Gamma}(1,1)\\
    t_{0}  &\sim \text{Uniform}(T_{0},T_{1}) ; \; \sigma^{2} \sim \text{Gamma}(1,1) ; \;
    Y_{t}  \sim \mathcal{N}( \alpha 2^{\beta(t-t_{0})}, \sigma^{2} )
\end{align*}
where $\alpha$, $\beta$, $t_{0}$, and $\sigma^{2}$ are parameters.
This model assumes that the expected number of cases doubles every $\frac{1}{\beta}$ time steps and never decreases. 
Though the choice of a Normal distribution assigns probability to values outside the total number of countries and/or cases, this non-zero probability is small.


\subsection{Forecast evaluation}

\subsubsection{Timeline}
We chose to evaluate 1-8 week ahead forecasts generated by humans and by computational models.
The number of weeks ahead that we could evaluate depended on which date individuals were asked to generate a forecast.
All questions asked for ground truth values on either July 1 or July 31.
If a question was associated with ground truth at a later date~(July 31) then we could evaluate longer forecast horizons~(Supp.~Table~\ref{table.comp_model_data_sources}).
In addition, we collected data on individual forecasters and studied these individual forecasts for the same time horizons.
For each question, computational models were given observed data from the data source used to determine the ground truth.
The ground truth was a single value on a given date, called the resolution date.

\subsubsection{Trend and Performance}
Trend and performance of forecasts were assessed using the median absolute error~(MAE) and weighted interval score~(WIS \cite{bracherEvaluatingEpidemicForecasts2021}). 
The MAE is presented to emphasize the trend in forecasts while WIS, which includes MAE, is a more formal evaluation metric.
Given a forecast~($F$) as a probability density function, the median absolute error is the median of $F$ minus ground truth.  
An individual forecaster's bias~($b$) is defined as their median prediction~($m$) minus the ground truth~($g$), or 
$
    b = m-g. 
$
By computing the bias for all individual forecasters, we can then compute sample statistics from this population of biases.
Because the ground truth for all six targets is positive, a negative $b$ indicates a prediction that is smaller than the ground truth and vice-versa.

The WIS takes as input a select number of quantiles from a forecast~$(F)$ and the truth $y$ and then returns a number from zero to infinity.
We chose to evaluate quantiles from the 1st percentile up to the 99th percentile in increments of 1 percent.
Smaller values of WIS indicate better performance, whereas larger WIS values indicate weaker performance. 
The WIS is an absolute score that depends on the magnitude of the target and so WIS scores cannot be compared across targets. 
We used the scoringUtils package in R to compute WIS~\cite{bosse2022evaluating}
To aid in the interpretation of WIS values, and to scale weighted interval scores for different targets in a unified way, they have been rescaled as relative WIS. 
We define the relative WIS as the weighted interval score for one forecast divided by the weighted interval score for the random walk presented above---a model that assumes no trend---for the same target at the same time horizon. 

\subsection{Forecaster rationales and interpretation}
Forecasters on the Metaculus platform had the option to attach to their forecasts a text-based rationale or comment. 
All text data is available online~(See Table~\ref{table.comp_model_data_sources} for links to questions).
Text data was evaluated and interpreted based on the author's content-area expertise.
The goal was to identify a small set of themes---concepts that were frequently discussed in the comments---from forecasters.
There was no formal qualitative analysis conducted.

\subsection{Data Availability}
Forecasting data, code to process forecasts, and code to generate figures in this manuscript are available at \url{https://github.com/computationalUncertaintyLab/mpox_eval}.
Real time forecasts about the mpox outbreak are available at \url{https://www.metaculus.com/project/monkeypox/}.  

\section{Results}

Our results may be categorized into seven broad findings, each of which are further delineated below.
We find that (1) as forecast horizon shrinks, human judgment predictions intervals do not show evidence of shrinking and that the human judgment ensemble~(HJE) often assigns positive probability to implausible values; (2) individual forecasters tended to predict values that were smaller than the ground truth; the HJE had (3a)~an MAE and WIS that was better than a random walk~(RW), autoregressive~(AR), and doubling time model~(DT) at forecast horizons of two weeks or greater, and 3b) WIS that was worse than a RW model at a one week ahead horizon;
(4)~well-performing forecasts may be in part due to the skew and large uncertainty introduced when soliciting forecasts on a logarithmic scale; 
(5)~the mode of individual human judgment forecasts was associated with the last observed value of the time series to be predicted; (6)~six major themes were present in text comments submitted by forecasters with the most frequent theme being `Taking cues from epidemiology and surveillance'; and  (7)~forecasters tended to make forecasts within two weeks of when a question was posed and continually revised their forecasts over time.

\subsection{Temporal trends in human judgement predictions}

For continuous targets, the median HJE prediction moved closer to the truth as the resolution date was approached, but the 80\% prediction interval for HJE did not show evidence of decreasing at the same rate as the resolution date was approached.~(Fig.~\ref{fig.hjandtruth}). 

For the binary question regarding the PHEIC, the median and the 80\% prediction interval were not appreciably changed as the resolution date was approached.
The median [80\% prediction interval] at three, two, and one weeks before the PHEIC declaration equalled 0.39 [0.15, 0.70],  0.39 [0.15, 0.67], and 0.40 [0.15, 0.70], respectively ~(Fig.~\ref{fig.hjandtruth}~bottom right).
     
The observed time series for continuous targets followed a non-negative trend over time.
This is because the questions posed asked for predictions of cumulative targets, i.e., cumulative number of mpox cases, countries with an mpox case, etc.
The HJE assigned probability to values well below the most recently reported true value at three, two, and one week ahead~(see Supp. Fig.~\ref{supp1.densities_and_truth}).   
For example, at one week before the ground truth, the HJE 10th percentile~(lower bound for a 90\% prediction interval) for the question `What will be the total number of confirmed and suspected monkeypox cases in Canada as of July 1, 2022?' was equal to 38 despite the number of true cases at that time point equaling 260.
The assignment of positive probably far below the reported true number of cases at the time of forecasting likely rules out issues with lag or revisions to the observed time series of cases. 

\subsection{Under prediction of outbreak}

The bias~(defined above in Methods) is negative for all targets at all forecast horizons before the ground truth is determined and moves closer to zero as the forecast horizon shrinks~(Fig.~\ref{fig.mediansovertime}).
We observed a positive bias for only one target for two-day ahead forecasts.
For predictions of the total number of cases that would be reported in Canada by July 01, 2022, the bias was negative until June 24th, close to zero between June 24th and June 29th, and positive from June 29th until the ground truth was observed.

An alternative way to characterize this bias is to compute the proportion of individual forecasts with a median prediction above the observed ground truth~(See Supplemental Fig.~\ref{supp.propssovertime}).

\subsection{Comparison of trend and performance between human judgment and computational models}

\begin{figure}
    \centering
    \includegraphics{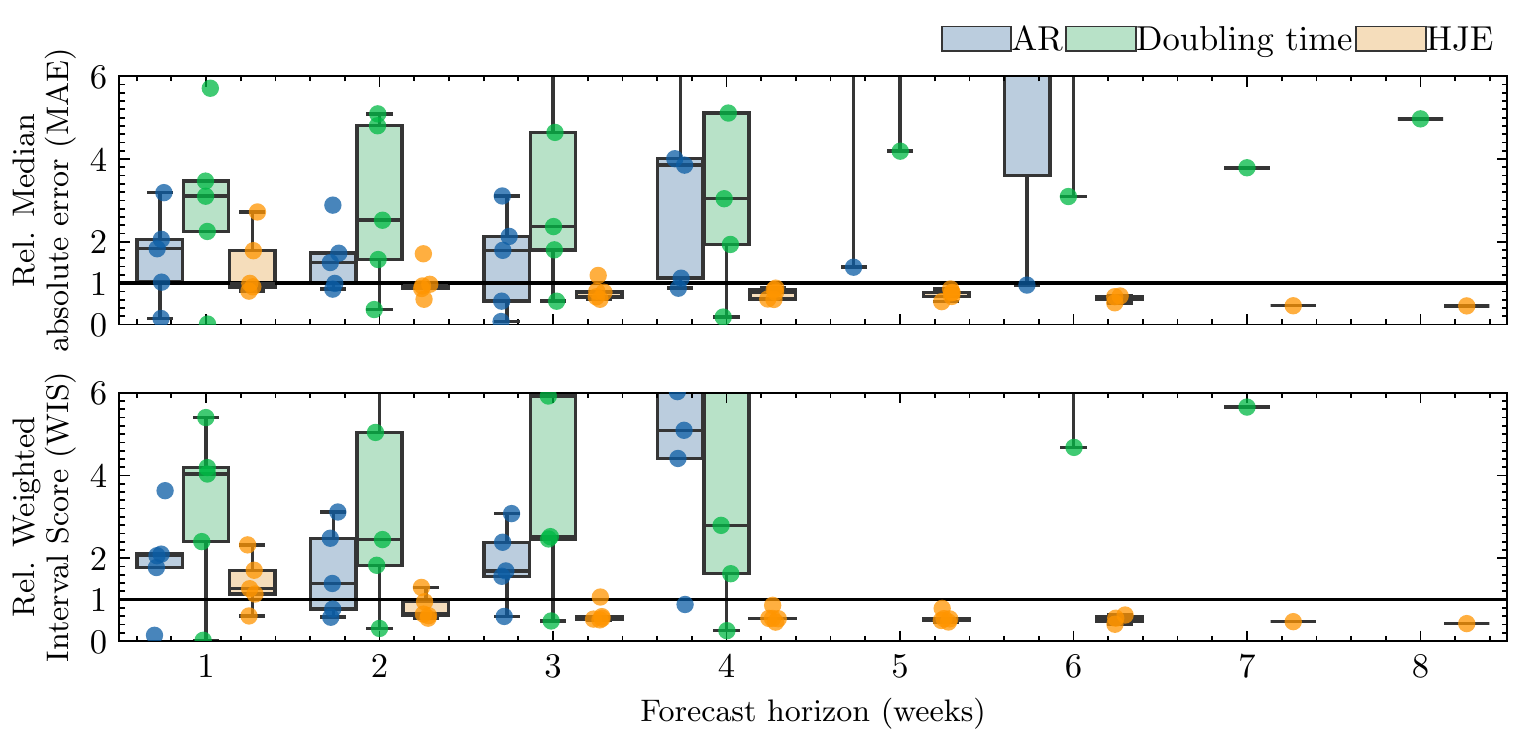}
    \caption{The (A.)~median absolute error~(MAE) and (B.)~weighted interval score~(WIS) for the auto-regressive model~(AR), doubling time, and human judgment ensemble~(HJE) relative to the random walk.
    MAE/WIS are plotted for each individual question and summarized with a boxplot.
    A horizontal line in black is plotted at the value 1~(i.e. a forecast that performs equally to a forecast generated from the random walk model).
    \label{fig.accuracy}}
\end{figure}

HJE had improved~(i.e.~smaller) median absolute error and improved~(i.e.~smaller) WIS compared to a random walk~(RW), auto-retrogressive~(AR), and doubling-time~(DT) model for forecast horizons of 2-8 weeks~(Fig.~\ref{fig.accuracy}).
For one week ahead forecasts, the median absolute error and WIS was better than AR and DT models but worse than the RW model.

\begin{figure}
    \centering
    \includegraphics{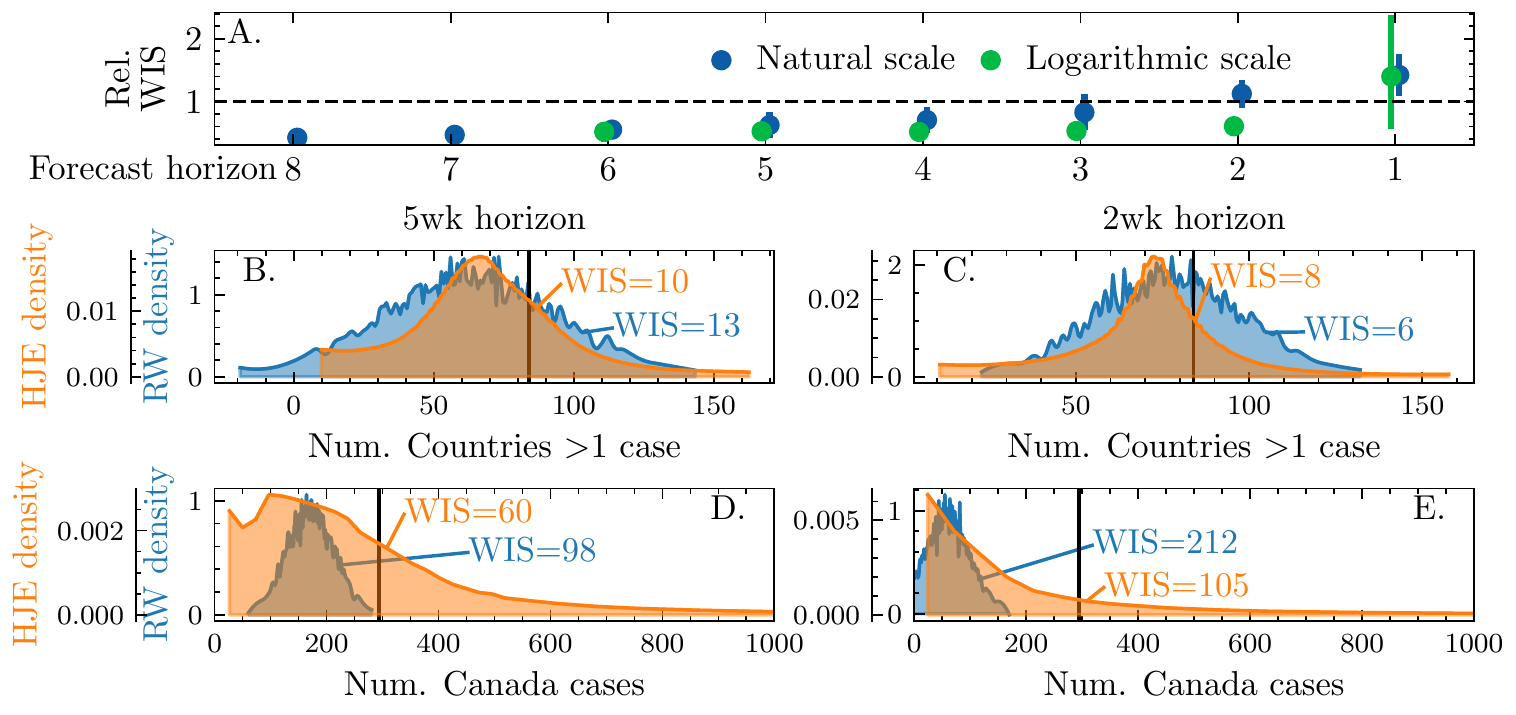}
    \caption{(A.)~Relative WIS between the human judgment ensemble~(HJE) and a random walk~(RW) over forecast horizons of 1-8 weeks and stratified by forecasts  solicited on a logarithmic~(green) or natural~(blue) scale. 
    Values less than 0 indicate human judgment ensemble forecasts performed better and values greater than 0 indicate the random walk model performed better.
    (B.)-(D.) Predictive densities for the human judgment ensemble~(orange) and random walk~(blue) at 5 week~(B. and D.) and 2 week~(C. and E.) forecast horizons for the number of countries to report at least one mpox case~(natural scale) and number of reported cases in Canada~(logarithmic scale)~\label{fig.bias_log}}
\end{figure}

Forecasts solicited on a logarithmic scale tended to perform better than forecasts solicited on natural scale~(Fig.~\ref{fig.bias_log}~A. and Table~\ref{tab.regtable}).
The difference in performance between forecasts solicited on logarithmic vs natural scale was most pronounced at 2 and 3 week ahead horizons.

When forecasting on a natural scale, human judgment forecasts naturally sharpen though not necessarily over the ground truth~(Fig.~\ref{fig.bias_log} B. and C.). 
When solicited on logarithmic scale, forecasts do not sharpen over the ground truth but the long tail includes ground truth~(Fig.~\ref{fig.bias_log}~D. and E.). 
The (possibly unintended) skew introduced by soliciting forecasts on a logarithmic scale may have improved WIS scores.

\subsection{Forecasts on logarithmic scale create immoderate skew}

\begin{figure}
    \centering
    \includegraphics{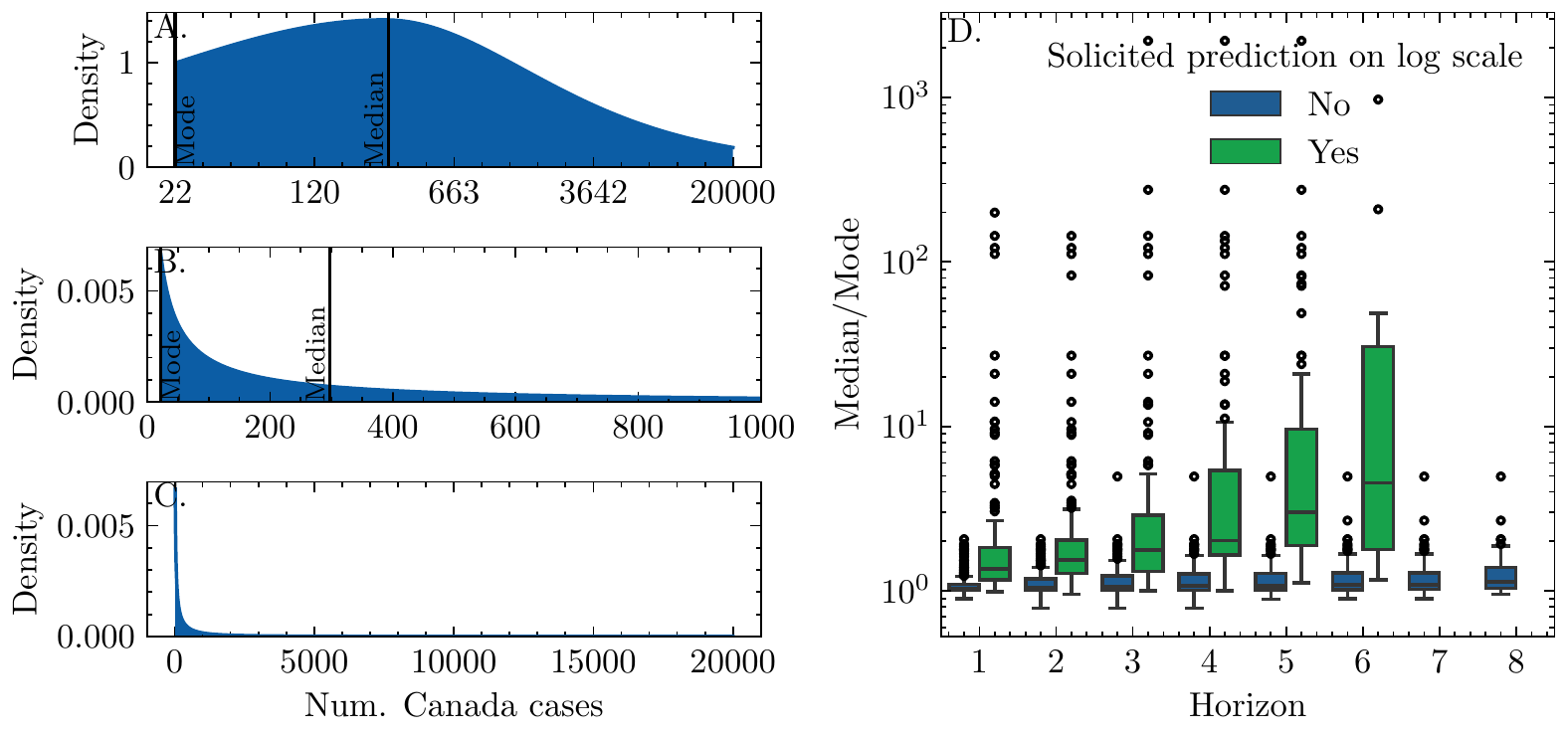}
    \caption{ A forecast for the number of mpox cases in Canada generated by a randomly selected single human 5 weeks before the ground truth:~(A.) on the logorithmic scale which was used by the Metaculus platform to capture forecasts, (B.) on the transformed natural scale for values smaller than 1000 cases, and (C.) the forecast on the natural scale over all possible values solicited on the platform. 
    (D.) The mode/median~(a measure of skew) for individual human judgment forecasts at forecast horizons 1-8, stratified by questions that solicited forecasts on a logarithmic scale~(green) and on natural scale~(blue). \label{fig.skew}}
\end{figure}

Soliciting forecasts on a logarithmic scale tended to right skew individual human judgment predictive densities
~(Fig.~\ref{fig.skew}). 
Forecasts generated on the logarithmic scale likely appear reasonable to the human eye~(Fig.~\ref{fig.skew}~A.); however, transforming this density back to the natural scale highlights the right-skew introduced~(Fig.~\ref{fig.skew}~B. and C.).
The skew---presented as the median divided by the mode for an individual human judgment forecast---on a logarithmic scale increased as forecast horizon increased. 
This is because humans tended to submit forecasts that are more uncertain at larger forecast horizons.
There is minimal skew for forecasts solicited on a natural scale, and this skew does not tend to change with increased forecast horizon.


The mode~(value with largest density value) of the predictive density for individuals was associated with the most recently observed value of the target at the time an individual produced a forecast~(Fig.~\ref{fig.last_obs}).
This association was present for 1-3 week ahead forecasts, 4-6 week ahead forecasts, and 7-8 week ahead forecasts~(see Table~\ref{tab.stats}).

\subsection{Themes present in the community that forecasted mpox}

We identified six potential themes from the comments associated with mpox forecasting targets: (1)~Taking cues from external data~(excluding surveillance data of the target itself) and news; (2)~introducing code or algorithms for how to produce a forecast; (3)~signaling data updates to the crowd; (4)~confusion about the target or ground truth; (5)~introspection about past forecasts; and (6)~taking cues from epidemiology or surveillance.

The most frequent~(N=6 comments) theme present in comments was discussion that linked to mpox epidemiology, such as symptoms exhibited by patients; the process by which the WHO declares a public health emergency of international concern; and how testing practices may impact how cases are reported~(Table~\ref{tab.jointdisplay} and see questions QID=10977, QID=10979 with urls in Supp.~\ref{table.comp_model_data_sources}).

Similarly, humans used cues from external datasets~(not the surveillance reports that would eventually contain the ground truth) and the news cycle to generate forecasts. 
Humans submitted URLs to news articles that they felt were related to the target of interest and included python code to aide others in keeping track of targets over time~(N=5; see QID=10975, 11039, 10977, 10978). 

It is important to note a frequent comment was related to confusion about how to define the ground truth and the final value of the ground truth~(N=4; QIDs=10975, 10978, 10979,10981).

\subsection{Frequency and timing of forecasts}

\begin{figure}[ht!]
    \centering
    \includegraphics[width=\textwidth]{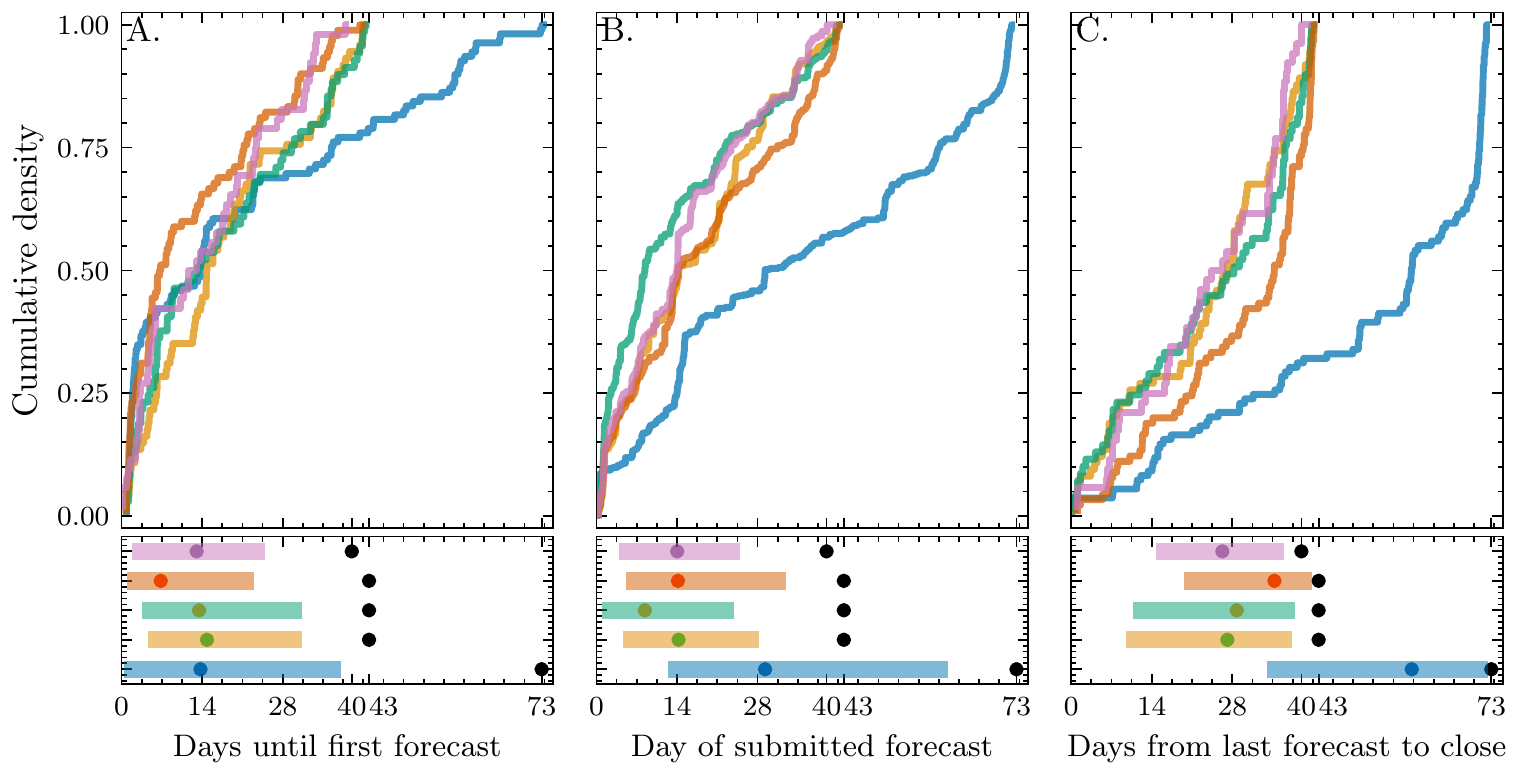}
    \caption{
    Empirical cumulative density function~(ecdf) of (A.)~the number of days from when a question was posed to the crowd to when an individual submitted their first forecast, (B.)~when an individual submitted any forecast, including revisions, and (C.)~when an individual submitted their last forecast. 
    Below the ecdfs are corresponding medians~(colored circles), 25th percentiles~(left side of bars) and 75th percentiles~(right side of bars). 
    Black circles represent the number of days from when the question was posed to when the question was closed for submission.
    \label{fig.revisions}}
\end{figure}

Forecasters most often submitted their first forecast two weeks after the six considered questions were posted on Metaculus, submitted most revisions within one week of their first forecast, and submitted their last forecast one week after their first forecast~(Fig.~\ref{fig.revisions}~A.-C.).
An exception is the question that asked individuals to forecast the number of European countries that would report one or more cases of mpox.
Similar to other questions, forecasters submitted their first forecast within two weeks; however, the median time at which forecasters submitted their last forecast for this question was eight weeks after the question was first posted, or approximately six weeks after their first forecast~(Fig.~\ref{fig.revisions}~C.). 
The mean number of revisions across all questions was $3.58~(95\text{CI}:[0,16.45])$

\section{Discussion}

Crowdsourced human judgments can produce rapid, moderately accurate forecasts up to eight weeks ahead of an emerging outbreak.
The mechanics of how humans build forecasts varies. 
We suppose that it is likely humans generated their forecasts by extracting cues: from the surveillance system used to evaluate the ground truth, news cycle, and from one another. 
In particular, this crowd was likely anchored to recent surveillance observations.
We hypothesize that the way in which forecasts were solicited has a large impact on the forecasts that were submitted~\cite{scapolo2006eliciting,hoffman2006method,o2019expert}, and, though soliciting forecasts on a logarithmic scale resulted in better WIS than the three computational models, this solicitation method may not be representative of an individual's intended forecast.

In particular, we would like to highlight several expected and unexpected properties of the human judgment ensemble forecasts presented for the six considered questions. 

\begin{figure}[ht!]
    \centering
    \includegraphics[width=\textwidth]{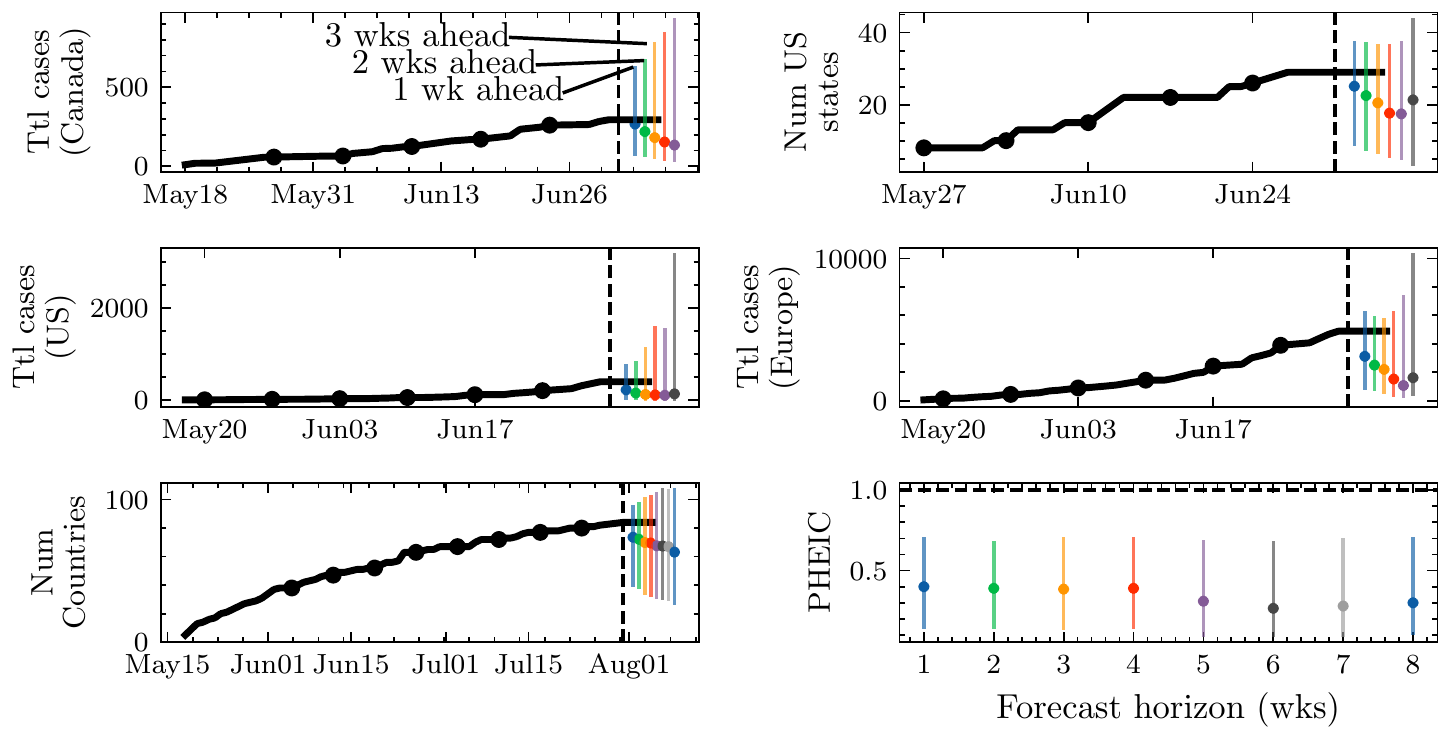}
    \caption{
    Human judgment ensemble forecasts for one week ahead and up to eight weeks ahead for six questions where the truth was collected.
    The dashed, black vertical is when the ground truth was determined.
    The one week ahead forecast is closest the dashed line, the two week ahead is second closest, etc.
    Black circles mark the observed values at one through eight weeks ahead.
    The 10th predictive quantile for human judgment is the bottom of the line, the 90th quantile is the top of the line, and the median (50th quantile) is presented as a circle.
    \label{fig.hjandtruth} }
\end{figure}

\begin{figure}
    \centering
    \includegraphics{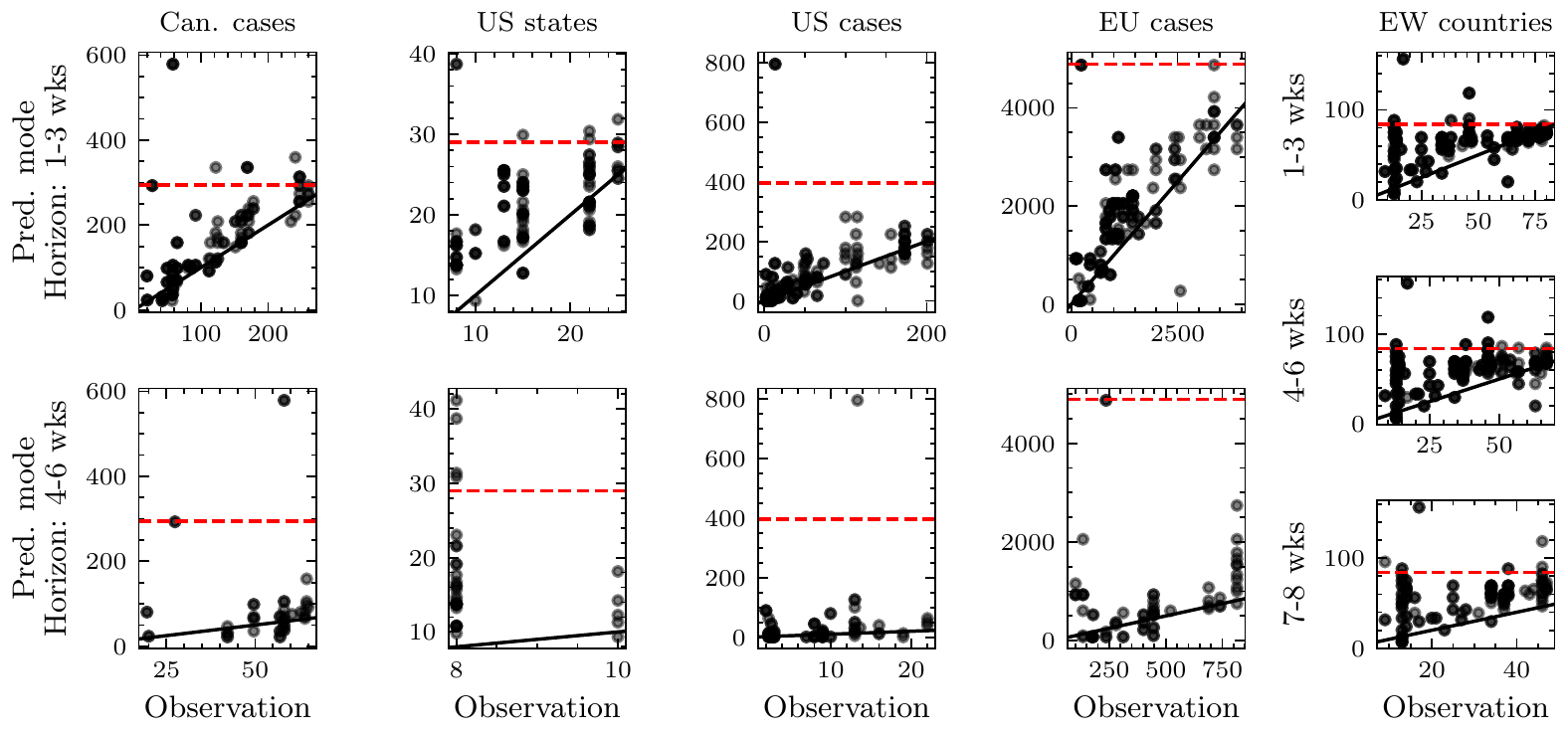}
    \caption{The observed value of the target quantity at the time of prediction~(horizonal axis) vs the mode of individual human judgment forecasts~(vertical axis) stratified by target and by forecast horizon. Red line denotes the final ground truth. Black line is the identity line. \label{fig.last_obs}}
\end{figure}

We expected, and observed, that as the resolution date was approached the difference between the median prediction and ground truth would decrease~(Fig.~\ref{fig.hjandtruth}).
This is in line with the intuition that more data~(via a longer time series) and less randomness~(shorter time interval for forecast) will produce a more accurate prediction.
Comments submitted by individuals support the claim that they continued to monitor surveillance reports about the time series that they were forecasting and collecting data about this time series as time passed~(Table.~\ref{tab.jointdisplay}).
For example, forecasters commented ``The total number of confirmed + suspected monkeypox cases in Europe on July 1st 2022 at 12pm EST as per this spreadsheet is 5762 cases. I have the HTML and a csv of the cases if anyone comes to a different number.'' and ``As of June 28, 27 states (not counting DC) have reported at least one case of monkeypox according to the U.S. CDC''.
In addition, the association between the most recently observed value and the mode for individual predictive densities is also supportive of a more accurate forecast as the resolution date was approached~(Fig.~\ref{fig.last_obs}).

\begin{figure}[ht!]
    \centering
    \includegraphics[width=\textwidth]{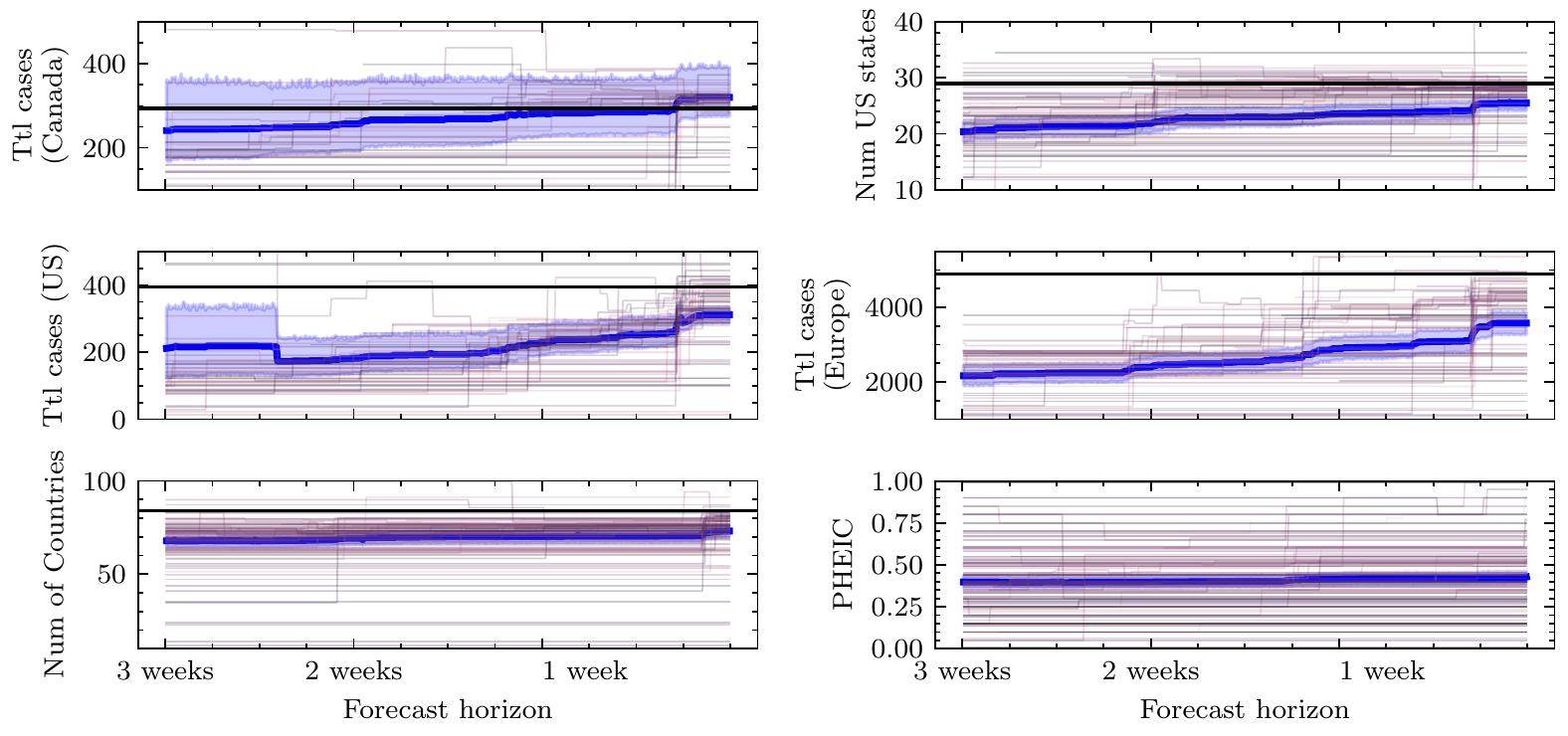}
    \caption{Individual median predictions~(gray lines) of  continuous targets from three weeks to one hour before the ground truth in intervals of one hour, and the average~(opaque blue line) plus 95\% confidence interval~(lighter blue shading) for the median prediction.  
    The ground truth is a solid black line.
    \label{fig.mediansovertime}}
\end{figure}

We also expected individuals to, on average, underestimate the size of the outbreak~(Fig.~\ref{fig.mediansovertime}).
This outbreak was largely driven by novel transmission pathways, and is to date several orders of magnitude larger than the largest prior outbreak of mpox~\cite{centers2003update}.
Similarly, in past work, experts routinely underestimated the number of cases reported during the COVID-19 pandemic in 2020~\cite{mcandrew2022expert}.

We expected the individual human judgment forecasts and the human judgment ensemble to assign probabilities to near-impossible events ~(See lower bound of prediction interval in Fig.~\ref{fig.hjandtruth}).
In past work, subject matter experts, those with experience forecasting, and laymen have often assigned positive probabilities to near-impossible events~\cite{mcandrew2022aggregating,mcandrew2022expert,recchia2021well}.
Comments from one individual forecaster of mpox highlighted this issue, commenting ``If my count is right, it's looking like 70 at the moment. About 35\% of the community weight is on impossible numbers.''.
We note that when submitting a forecast to Metaculus, individuals are not able to define the support~(i.e. the interval where there exists a non-zero probability) for their forecast, nor are they able to assign probability zero to any intervals.
This means that even if a forecaster knows that a given interval has a probability of zero the platform doesn't allow them to assign a zero.
To characterize the mechanism by which humans assign positive probabilities to impossible events, we suggest a study that conducts key informant interviews with forecasters who do or do not assign probability to near-impossible events.
For example, one mechanism could be that individuals may have expected the most recent value for an mpox target could be revised to be smaller than the currently reported value.
Even more, we recommend studying the difference in forecast performance between two platforms: one that does not allow the assignment of zero probability and one that does allow this type of assignment.

We did not expect that the uncertainty around these forecasts would be similar or not appreciably decrease as the resolution date was approached.
We posit that the most likely mechanism by which these intervals remained large was soliciting forecasts on a logarithmic scale~(Fig.~\ref{fig.skew}).
Forecasts solicited on a natural scale tended to become more focused around a single value, but forecasts solicited on a logarithmic scale tended to look exponential as the resolution date was approached~(Fig.~\ref{fig.bias_log}).
We cannot determine whether the humans for this forecasting challenge were susceptible to misunderstanding the difference between their forecast on a logarithmic vs. natural scale. 
A rigorous, randomized design that posed the same questions on both logarithmic and natural scales could provide evidence for this line of inquiry.
However, past work indicates that humans have a poor understanding of exponentials and logarithms~\cite{melnik2023my,wagenaar1975misperception,heckler2013student}.
That said, the large uncertainty produced by logarithmic solicitation `saved' forecasts by shifting probability density from small values~(where the ground truth was not) to much larger values~(where the ground truth was).
This is illustrated in Figure~\ref{fig.bias_log} panels D and E, comparing the random walk to human judgment ensemble forecast.
However, this positive gain in performance may have come at the cost of misrepresenting the true forecast that individual's intended to produce.


A major limitation to this work was the small number of targets that we could evaluate.
We expect future work to evaluate human judgment forecasts of the trajectory of an infectious agent on a larger set of targets.
We are also limited by not surveying forecasters about the process that they used to generate their forecasts.
Because we did not survey participants, we can only speculate about the cues that they relied on from submitted comments and try to infer mechanisms from collected forecasts.

Nevertheless, the forecasts we analyze here involve predictions for a novel outbreak, and
a number of the issues raised in this work are related to past issues raised in human judgment forecasting of other novel outbreaks~\cite{mcandrew2022expert}.
Beyond these well-accepted issues, we also bring attention to general issues with humans' misperceptions of logarithms, which require further investigation within the setting of human judgment modeling~\cite{melnik2023my,wagenaar1975misperception,heckler2013student}.
Moreover, we highlight technical challenges that human judgment platforms should aim to overcome.
One concrete recommendation is to allow individuals to first propose the interval where they will assign a positive probability before specifying a functional form over this interval.
This technical change, which would need to be developed for the current platform, could guard against humans assigning probability to unlikely values and improve forecast performance.

We feel human judgment has the potential to play a role in supporting public health efforts during an outbreak such as mpox, but only after intense efforts to address issues like the ones raised in this work.

\clearpage

\clearpage
\section{Figures}

\clearpage
\section{Tables}

\begin{table}[ht!]
    \centering
    \begin{tabular}{r|ccc}
        \hline
     Parameter & $\hat{\beta}$ & 95CI & p  \\
     \hline
      Intercept                &  0.42 & [-0.07,0.77] & 0.02\\
      Log scale (vs Nat scale) &  -0.21 & [ -0.48,0.08]  & 0.15\\
      Forecast horizon         & -0.16 & [-0.24,-0.08] & 0.01\\
         \hline
    \end{tabular}
    \caption{Linear regression fit to WIS~(dependent variable), whether the forecast was solicited on a logarithmic~(vs natural) scale, and forecast horizon.
    We present estimated coefficients~($\hat{\beta}$), 95\% confidence intervals~(95CI), and associated pvalues~(p)~\label{tab.regtable}}
\end{table}

\begin{table}[]
    \centering
    \begin{tabular}{lcr}
    \hline
    Question/Target              & Spearmens rho & p \\ 
    \hline
     Number of Cases Canada      \\
    \hspace{2mm} 1-3 weeks ahead & 0.72 & $<0.01$ \\
    \hspace{2mm} 4-6 weeks ahead & 0.43 & $<0.01$ \\

     Number of US states      \\
    \hspace{2mm} 1-3 weeks ahead & 0.53  & $<0.01$ \\
    \hspace{2mm} 4-6 weeks ahead & -0.23 & 0.20 \\

     Number of US cases      \\
    \hspace{2mm} 1-3 weeks ahead & 0.73 & $<0.01$\\
    \hspace{2mm} 4-6 weeks ahead & 0.40 & $<0.01$\\
    
    Number of Europe cases      \\
    \hspace{2mm} 1-3 weeks ahead & 0.76 & $<0.01$\\
    \hspace{2mm} 4-6 weeks ahead & 0.34 & $<0.01$\\

     Number of European countires      \\
    \hspace{2mm} 1-3 weeks ahead & 0.41 & $<0.01$\\
    \hspace{2mm} 4-6 weeks ahead & 0.29 & $<0.01$ \\
    \hspace{2mm} 7-8 weeks ahead & 0.27 & $<0.01$ \\
    \hline
    \end{tabular}
    \caption{Spearman's correlation and associated pvalue to assess the association between the mode of individual forecasts and the most recently observed value at the time of the forecast.\label{tab.stats}}
\end{table}

\begin{table}[ht!]
    \centering
    \begin{tabular}{lL{2.5in}L{3.0in}}
        \hline
      Frequency  & Theme & Example\\
      \hline
      5          & Taking cues from external data and news  & BNO: NEW: Washington state reports 1st case of monkeypox\\ 
      \hline
      2          & Introducing code or algorithms for how to produce a forecast & Since this value [total number of reported suspected+confirmed cases on July 1, 2022 at 12pm EST according to this spreadsheet that is being actively updated by Global.health] changes throughout the day, I am including some code now, so that come 12 pm EST, another person can verify the \# of confirmed + suspected cases in Europe.\\ 
      \hline
      5          & Signaling to the crowd data updates  & Based on the spreadsheet, 45 countries with suspected+confirmed cases so far.\\
      \hline
      4          & Confusion about the target or ground truth & Good Morning. Is 82 the correct resolution? Thanks\\ 
      \hline
      2          & Introspection about past forecast & Just as post-mortem now that the question is resolved, I want to say that I still stand behind [sic] my reasoning[...]\\ 
      \hline
      6          & Taking cues from epidemiology or surveillance & Interesting context on the vote taken by the WHO committee in which they did not recommend declaration of a PHEIC:\\ 
      \hline
    \end{tabular}
    \caption{The most common themes among comments submitted by forecasters.  \label{tab.jointdisplay}}
\end{table}


\clearpage
\appendix

\section{Supplemental}

\subsection{Forecast submission}
A forecaster on the Metaculus platform can submit a predictive density $f$ as a convex combination of up to five logistic distributions
    \begin{align}
        f(x) &= \sum_{i=1}^{5} \pi_{k} \, g(x | \mu_{k}, s_{k}) \\
        g(x) &= \frac{\exp \l(-(x-\mu)/s\r)}{ s(1+ \exp\l(-(x-\mu)/s\r) )} 
    \end{align} 
where $\pi_k$ is a non-negative weight associated with the $k$th logistic distribution $g(x | \mu_{k}, s_{k})$ and all weights are constrained to sum to unity.
By default, a forecaster is presented with a single logistic density with a slider bar underneath this density that contains a square and two circles~(see supplemental figure~\ref{supp.platform}). 
A forecaster can change the value of $\mu$ by shifting the square to the left for lower values and to the right for higher values. 
The parameter $s$ is adjusted by sliding the circles left and right independently. 
If they wish, a forecaster may press “add a component” to add additional logistic distributions (up to 5 component distributions in total).
A second slider will appear under the first and two additional sliders that control the weights ($\pi$) associated with the first and second logistic distribution. The predictive density a user creates is continually updated in the browser to facilitate iteratively building their forecast for submission.They are also shown the Community Forecast [needs to be defined] along with a Current Points calculation reflecting the points they may received based on the resolution (see \url{https://www.metaculus.com/help/faq/#howscore} for more details). 
When a forecaster is satisfied with their prediction, they press the submit button.
After the first submission, a forecaster may revise their original prediction as many times as they choose until the question is closed on the platform or the ground truth is reported.
The ensemble was hidden from individuals for X days or until there were X forecasts, whichever came first.

Forecasters construct their predictive density over a bounded interval, presented either on a linear scale or on a log scale (base 10). The choice of intervals, and linear vs. log scale, was made by forecast question developers at Metaculus.
After 10 predictions from 10 unique forecasters were submitted to the Metaculus platform, an ensemble of these predictions was revealed to the community.
In addition to a formatted method for building a forecast, users were also able to post comments that could be viewed by all other users of the Metaculus system.

\subsection{Ensemble building}
\begin{figure}[ht!]
    \centering
    \includegraphics[width=\textwidth]{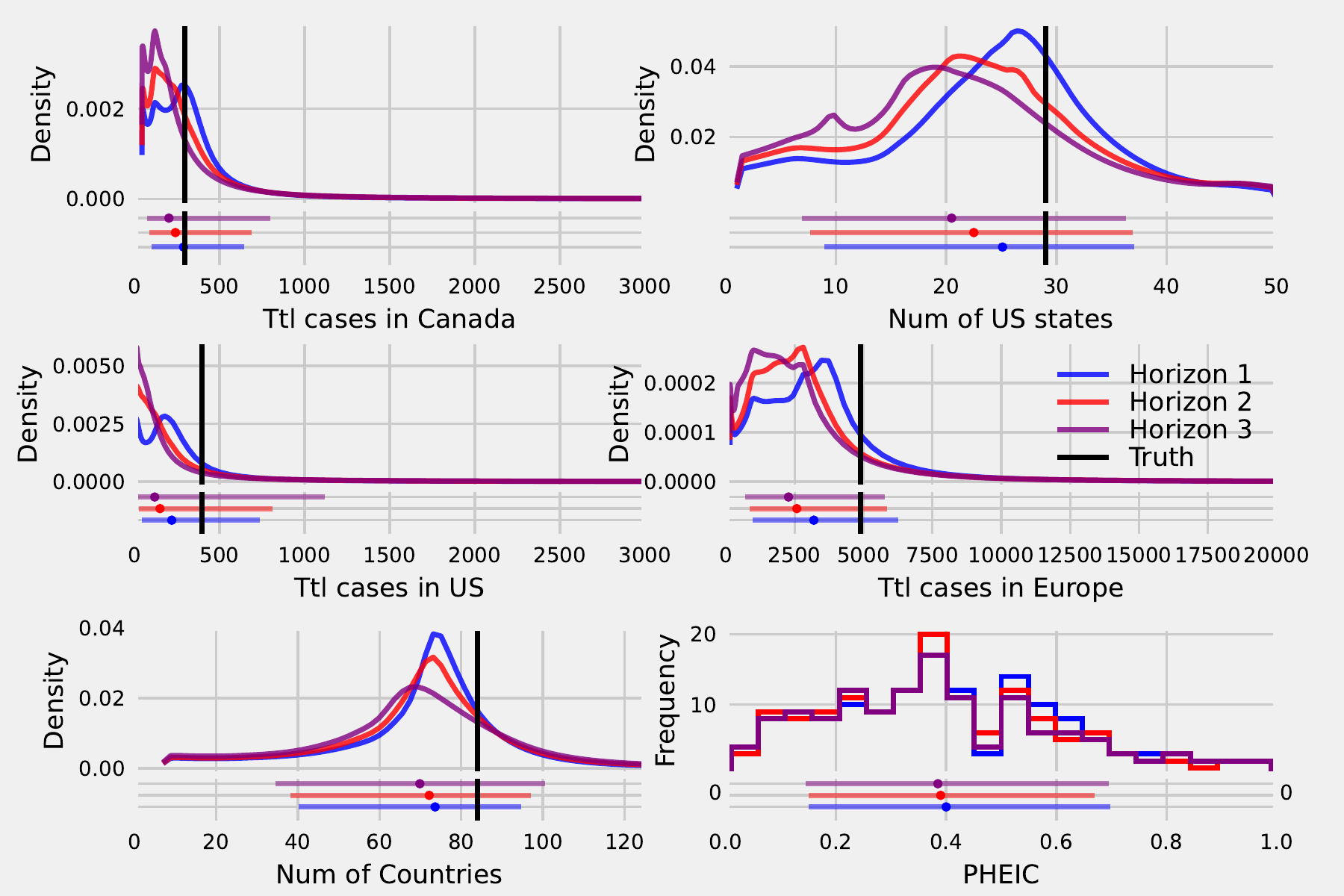}
    \caption{Human judgment ensemble predictive densities one (blue), two (red), and three (purple) weeks ahead of the date of the true value (black vertical line). Below each plot is presented a 25th quantile (leftmost point of the line), 50th quantile (dot), and 75th quantile (rightmost point of the line). 
    \label{supp1.densities_and_truth}}
\end{figure}

An ensemble predictive density $f$ in response to a question posed on the Metaculus platform was created as a weighted combination of all $M$ individual forecaster densities 
\begin{align}
    f(x) = \sum_{m=1}^{M} \pi_{m} f_{m}(x)
\end{align}  
where $f_{m}$ is a predictive density submitted by user $m$ and the weights assigned to forecaster $m$ are annotated as $\pi_{m}$. 
Weights could be a function of the forecaster’s accuracy on past questions and of how recently they submitted a prediction, but in this study we chose to weight all forecasts equally~(i.e., we assigned $\pi_{m} = 1/M \; \forall m$). 
The above ensemble strategy is used for questions where a forecaster submits a predictive density over a closed interval~(a continuum).
For questions entailing answers in the form of a scalar number, such as when forecasters are asked to submit a single probability of a particular event (see question (vi) above), the ensemble strategy build an empirical cumulative density function from predicted probabilities~(See Supp.~Fig.~\ref{supp1.densities_and_truth} bottom right panel for an example).

\begin{table}[ht!]
    \centering
    \begin{tabular}{lp{0.45\textwidth}ll}
       \hline
       Question &  Link to Metaculus question & Data source & Scale\\
               \hline
        Total cases in Europe & \url{https://www.metaculus.com/questions/10978/total-monkeypox-cases-in-europe-july-1-2022/} & Global health & Logarithmic \\ 
        Total cases in US      & \url{https://www.metaculus.com/questions/10979/total-monkeypox-cases-in-usa-on-july-1-2022/} & Global health & Logarithmic \\ 
        Total cases in Canada  & \url{https://www.metaculus.com/questions/11039/monkeypox-cases-in-canada-on-july-1-2022/} & Global health & Logarithmic \\ 
        Countries with at least one case & \url{https://www.metaculus.com/questions/10975/countries-with-monkeypox-by-july-31-2022/}  & Global health & Natural \\
        Number of US states              & \url{https://www.metaculus.com/questions/10981/-us-states-with-monkeypox-cases-on-july-1/}  & CDC & Natural \\
        WHO declaration of PHEIC         & \url{https://www.metaculus.com/questions/10977/who-declares-that-monkeypox-is-a-pheic/}       & WHO & Natural \\ 
        \hline
    \end{tabular}
    \caption{ Listing of data sources, forecast horizon cut points, and the data sources that were used as resolution criteria for questions posed to users on Metaculus. All datasets started  \label{table.comp_model_data_sources}}
\end{table}

These questions were:  (i) what will be the total number of confirmed and suspected monkeypox cases in Europe as of July 1, 2022?, (ii) what will be the total number of confirmed and suspected monkeypox cases in the United States as of July 1, 2022?, (iii) what will be the total number of confirmed and suspected monkeypox cases in Canada as of July 1, 2022? (iv) how many countries will confirm at least one case of monkeypox by July 31, 2022?, (v) how many states in the United States will report having a case of monkeypox as of July 1, 2022?, and (vi) will the World Health Organization declare the spread of monkeypox a Public Health Emergency of International Concern before 2023? We received a total of 1275 probabilistic predictions (initial forecasts or subsequent updates) from 442 individual forecasters. 

\begin{figure}[ht!]
    \centering
    \includegraphics[width=\textwidth]{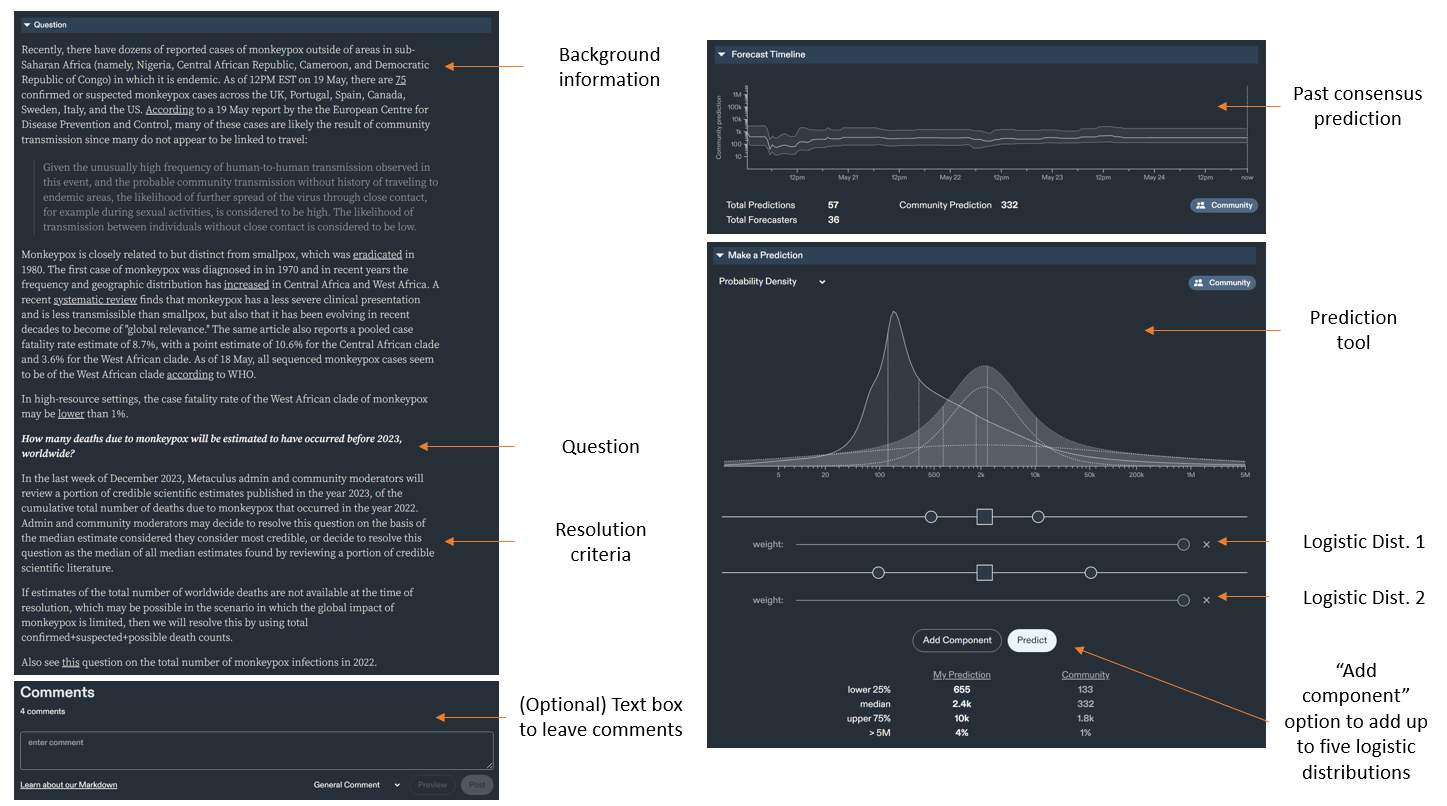}
    \caption{Example of the Metaculus platform used by forecasters to generate probabilistic predictions of questions posed about monkeypox.\label{supp.platform}}
\end{figure}

\clearpage

\section{Figures}
\appendix

\begin{figure}[ht!]
    \centering
    \includegraphics[width=\textwidth]{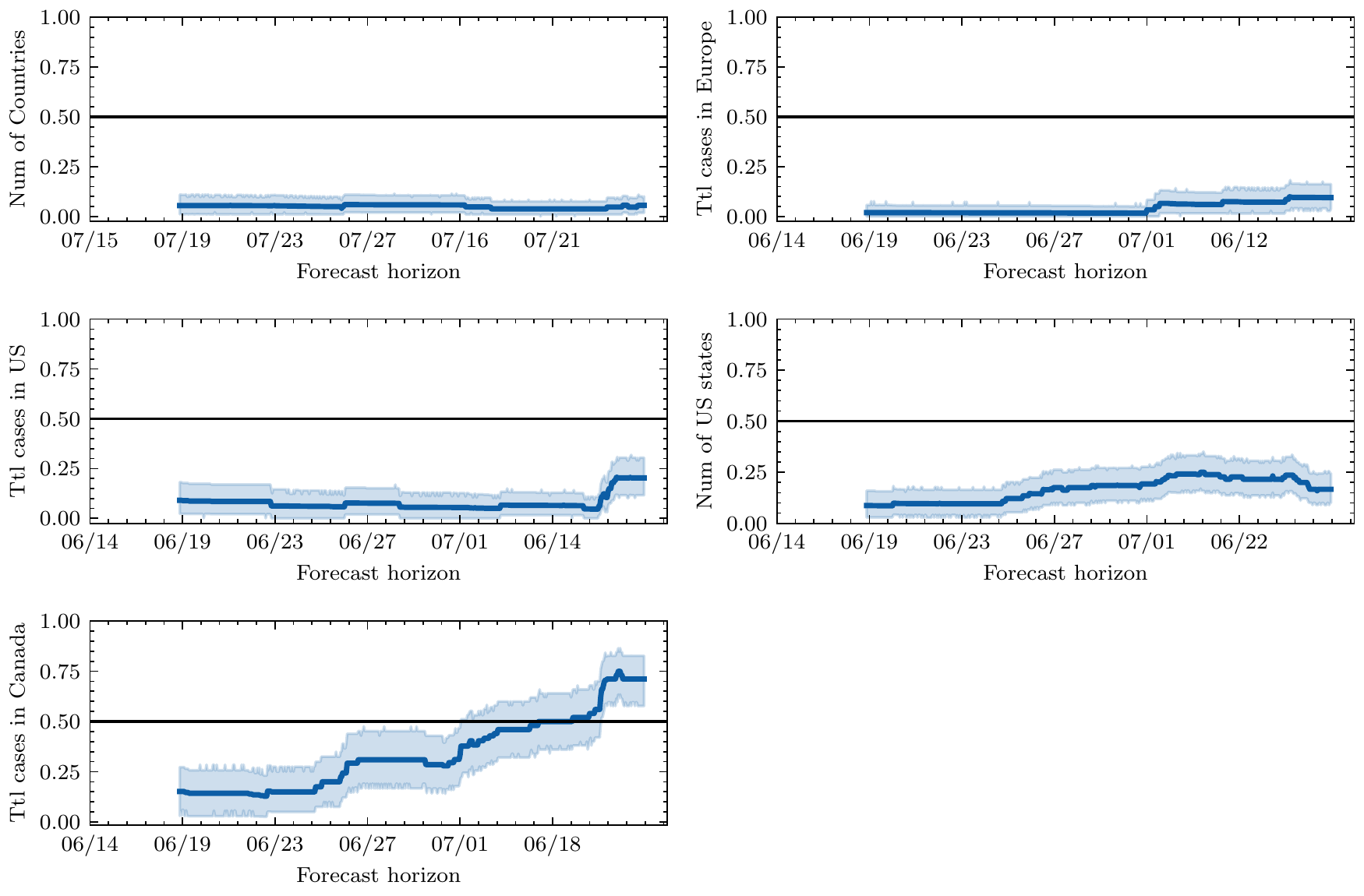}
    \caption{The proportion plus 95\% confidence intervals of individual median predictions that were greater than the ground truth for five continuous targets from three weeks before the ground truth to one week before the ground truth in intervals of one hour.
    The majority of human judgment forecasters submitted densities with a median below the ground truth. 
    Only when asked to predict the total number of cases in Canada did human forecasters over estimate the severity of the mpox outbreak.\label{supp.propssovertime}}
\end{figure}


\end{document}